\documentclass[]{pasj01}
\usepackage{bmpsize}
\usepackage{ulem}

\jyear{2019}

\Received{$\langle$reception date$\rangle$}
\Accepted{$\langle$accept date$\rangle$}
\Published{$\langle$publication date$\rangle$}

\begin{document}

\title{The Hyper Suprime-Cam SSP Transient Survey in COSMOS: Overview}
\author{
Naoki \textsc{Yasuda}\altaffilmark{1,*},
Masaomi \textsc{Tanaka}\altaffilmark{2,1},
Nozomu \textsc{Tominaga}\altaffilmark{3,1}, 
Ji-an \textsc{Jiang}\altaffilmark{4}, 
Takashi J. \textsc{Moriya}\altaffilmark{5},
Tomoki \textsc{Morokuma}\altaffilmark{4,1},
Nao \textsc{Suzuki}\altaffilmark{1}, 
Ichiro \textsc{Takahashi}\altaffilmark{1,6},
Masaki S. \textsc{Yamaguchi}\altaffilmark{3},
Keiichi \textsc{Maeda}\altaffilmark{7,1}, 
Masao \textsc{Sako}\altaffilmark{8},
Shiro \textsc{Ikeda}\altaffilmark{9,1},
Akisato \textsc{Kimura}\altaffilmark{10}, 
Mikio \textsc{Morii}\altaffilmark{9},
Naonori \textsc{Ueda}\altaffilmark{10}, 
Naoki \textsc{Yoshida}\altaffilmark{11,1,6},
Chien-Hsiu \textsc{Lee}\altaffilmark{12},
Sherry H. \textsc{Suyu}\altaffilmark{13,14},
Yutaka \textsc{Komiyama}\altaffilmark{5,15},
Nicolas \textsc{Regnault}\altaffilmark{16}, and
David \textsc{Rubin}\altaffilmark{17,18}
}%
\altaffiltext{1}{Kavli Institute for the Physics and Mathematics of the Universe (WPI), The University of Tokyo Institutes for Advanced Study, The University of Tokyo, 5-1-5 Kashiwanoha, Kashiwa, Chiba 277-8583, Japan}
\altaffiltext{2}{Astronomical Institute, Tohoku University, Aoba, Sendai 980-8578, Japan}
\altaffiltext{3}{Department of Physics, Faculty of Science and Engineering, Konan University, 8-9-1 Okamoto, Kobe, Hyogo 658-8501, Japan}
\altaffiltext{4}{Institute of Astronomy, Graduate School of Science, The University of Tokyo, 2-21-1 Osawa, Mitaka, Tokyo 181-0015, Japan}
\altaffiltext{5}{National Astronomical Observatory of Japan, National Institutes of Natural Sciences, 2-21-1 Osawa, Mitaka, Tokyo 181-8588, Japan}
\altaffiltext{6}{CREST, JST, 4-1-8 Honcho, Kawaguchi, Saitama 332-0012, Japan}
\altaffiltext{7}{Department of Astronomy, Kyoto University, Kitashirakawa-Oiwake-cho, Sakyo-ku, Kyoto 606-8502, Japan}
\altaffiltext{8}{Department of Physics and Astronomy, University of Pennsylvania, Philadelphia, PA 19104, USA}
\altaffiltext{9}{Research Center for Statistical Machine Learning, The Institute of Statistical Mathematics, 10-3 Midori-cho, Tachikawa, Tokyo 190-8562, Japan}
\altaffiltext{10}{NTT Communication Science Laboratories, 2-4 Hikaridai, Seika-cho, Keihanna Science City, Kyoto 619-0237, Japan}
\altaffiltext{11}{Department of Physics, Graduate School of Science, The University of Tokyo, 7-3-1 Hongo, Bunkyo-ku, Tokyo 113-0033, Japan}
\altaffiltext{12}{Subaru Telescope, National Astronomical Observatory of Japan, 650 N A’ohoku Pl., Hilo, HI 96720, USA}
\altaffiltext{13}{Max-Planck-Institut f\"{u}r Astrophysik, Karl-Schwarzschild-Str. 1, 85748 Garching, Germany}
\altaffiltext{14}{Physik-Department, Technische Universit\"{a}t M\"{u}nchen, James-Franck-Stra\ss e 1, 85748 Garching, Germany}
\altaffiltext{15}{Graduate University for Advanced Studies (SOKENDAI), 2-21-1 Osawa, Mitaka, Tokyo 181-8588, Japan}
\altaffiltext{16}{Sorbonne Universit\'e, Universit\'e Paris Diderot, CNRS/IN2P3, Laboratoire de Physique Nucl\'eaire et de Hautes Energies, LPNHE, 4 Place Jussieu, 75252 Paris, France}
\altaffiltext{17}{Physics Division, Lawrence Berkeley National Laboratory, 1 Cyclotron Road, Berkeley, CA 94720, USA}
\altaffiltext{18}{Space Telescope Science Institute, 3700 San Martin Drive, Baltimore, MD 21218, USA}

\email{naoki.yasuda@ipmu.jp}


\maketitle

\begin{abstract}
We present an overview of a deep transient survey of the COSMOS field with the Subaru Hyper Suprime-Cam (HSC). The survey was performed for the 1.77 deg$^2$ ultra-deep layer and 5.78 deg$^2$ deep layer in the Subaru Strategic Program over 6- and 4-month periods from 2016 to 2017, respectively. The ultra-deep layer shows a median depth per epoch of 26.4, 26.3, 26.0, 25.6, and 24.6 mag in $g$, $r$, $i$, $z$, and $y$ bands, respectively; the deep layer is $\sim0.6$ mag shallower. In total, 1,824 supernova candidates were identified. Based on light curve fitting and derived light curve shape parameter, we classified 433 objects as Type Ia supernovae (SNe); among these candidates, 129 objects have spectroscopic or COSMOS2015 photometric redshifts and 58 objects are located at $z > 1$. Our unique dataset doubles the number of Type Ia SNe at $z > 1$ and enables various time-domain analyses of Type II SNe, high redshift superluminous SNe, variable stars, and active galactic nuclei.
\end{abstract}

\section{Introduction}
Time-domain astronomy, now a major field of astronomy, continues to reveal fascinating variable and transient phenomena in the universe. Inspired by the discovery of dark energy through Type Ia supernovae (SNe Ia, \cite{riess98a,perlmutter99a}), many large and systematic surveys have been conducted over the last decade to test the $\Lambda$ cold dark matter (CDM) model with high precision \citep{astier06a,frieman08a,freedman09b}.   
These surveys demonstrate that the time-domain dataset is very rich, providing information on SN~Ia cosmology and other recent discoveries, such as the ``super-Chandrasekhar'' supernova (SN) \citep{howell06a}, superluminous supernova (SLSN, \cite{quimby07b}), gravitationally lensed SN \citep{goobar17a}, and rapidly evolving transients \citep{drout14,pursiainen18a}.

For SN~Ia cosmology, the latest Panoramic Survey Telescope and Rapid Response System 1 (Pan-STARRS1) survey ($z < 0.6$) reports that the cumulative number of spectroscopically confirmed SNe~Ia is now 1,049 \citep{scolnic17a}, and the ongoing Dark Energy Survey \citep{bernstein2012, abbott16b} ($z < 1.0$) is about to add a few thousand SNe~Ia to the Hubble Diagram \citep{abbott18a}. However, the number of SNe~Ia in high redshift ($z > 1.0$) is still limited since the deep surveys have been possible only with the Hubble Space Telescope (HST), whose field of view is very small \citep{suzuki12a,riess18a}. We aim to probe the high redshift universe and trace the expansion history of the universe from the deceleration epoch through the acceleration epoch, to determine whether dark energy is time-variable or not. In this paper, we describe our transient survey using the 8-m Subaru telescope and highlight the expected scientific outcomes.

\begin{figure*}
\begin{center}
\includegraphics[width=2\columnwidth]{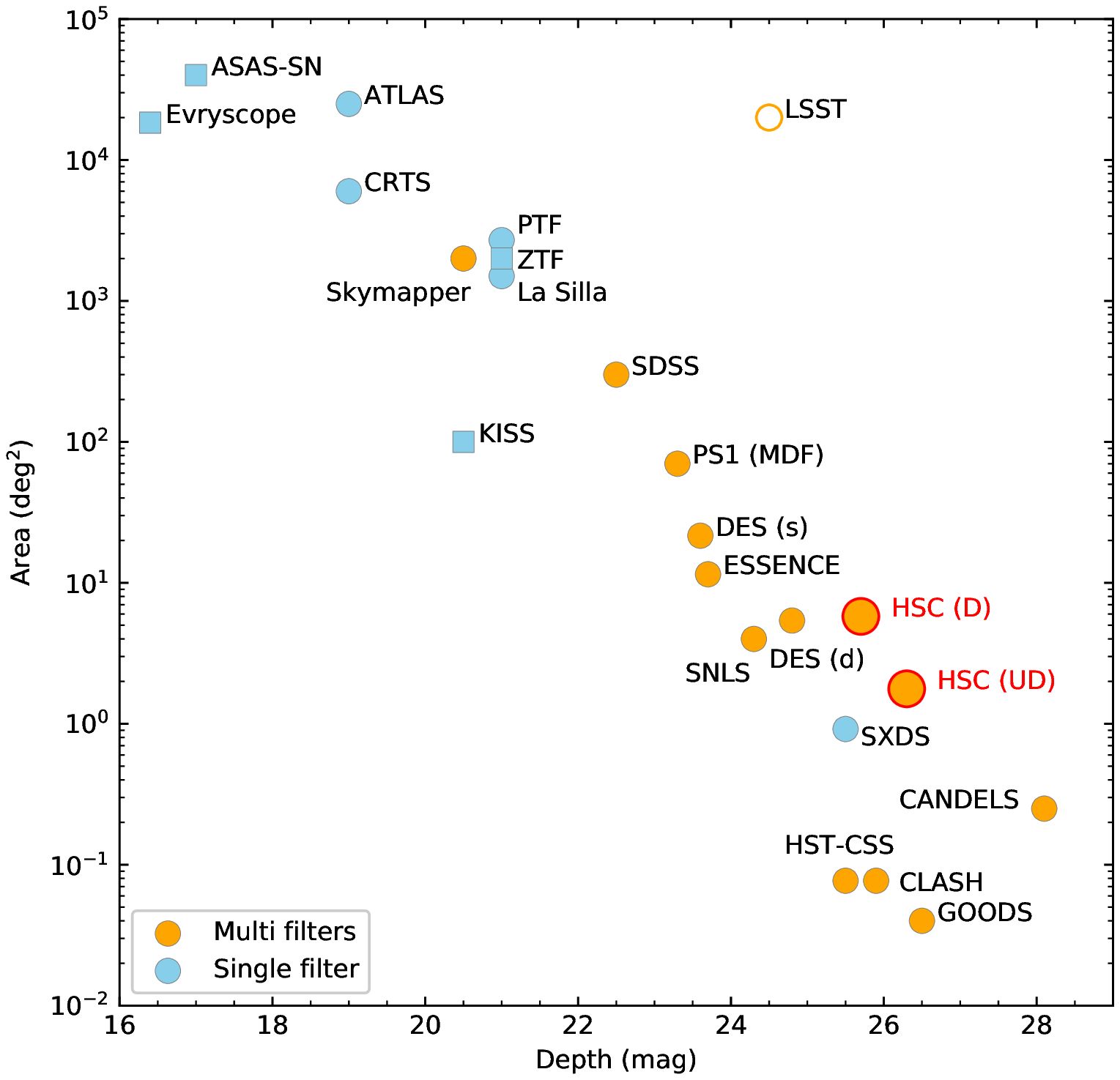}
\end{center}
\caption{Summary of typical survey depth (in the optical band) and area for long-term transient surveys: All-Sky Automated Survey for Supernovae (ASAS-SN, \cite{kochanek17,holoien17}); Asteroid Terrestrial-impact Last Alert System (ATLAS, \cite{tonry18}); Evryscope \citep{law15}; Catalina Real-Time Transient Survey (CRTS, \cite{drake09,djorgovski11}); Palomar Transient Factory (PTF, \cite{rau09a,law09a}); Zwicky Transient Facility (ZTF, \cite{bellm19}); Kiso Supernova Survey (KISS, \cite{morokuma14}); Skymapper \citep{keller07,scalzo17}; La Silla-QUEST Low Redshift Supernova Survey \citep{baltay13}; Sloan Digital Sky Survey (SDSS, \cite{frieman08a}); Pan-STARRS1 (PS1, \cite{rest14}); Supernova Legacy Survey (SNLS, \cite{astier06a}); ESSENCE \citep{miknaitis07}; Dark Energy Survey (DES, \cite{dandrea18}); Subaru/XMM-Newton Deep Survey (SXDS, \cite{morokuma08}); Hubble Space Telescope Cluster Supernova Survey (HST-CSS, \cite{Dawson2009}) HST/GOODS \citep{dahlen12}; HST/CANDELS \citep{rodney14}; and HST/CLASH \citep{postman12,graur14}.
Orange and blue points show multi-filter and single-filter surveys, respectively. Surveys shown with a square symbol indicate high-cadence surveys ($<1$ day).
}
\label{fig:survey}
\end{figure*}

\begin{figure*}
\begin{center}
\includegraphics[width=2\columnwidth]{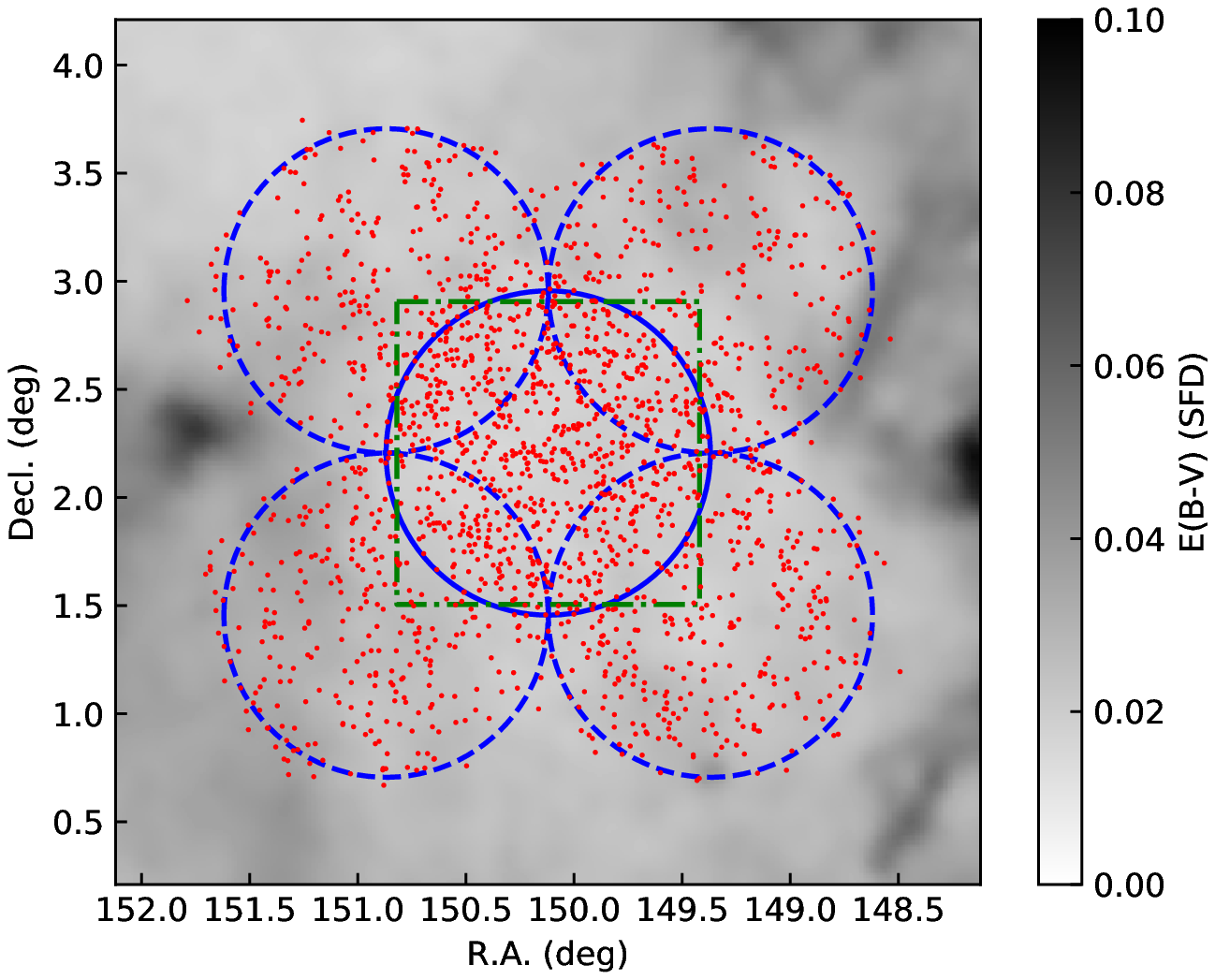}
\end{center}
\caption{Pointing layout on the sky (ultra-deep: blue (solid), deep: blue (dashed), original COSMOS \citep{scoville2007cosmos} coverage: green (dash dot)) overlaid on an SFD 
\citep{SFD} reddening map. Positions of detected supernova (SN) candidates are indicated by red points. Given that we were dithering around fiducial pointings, the actual coverage is wider than that indicated by the dashed blue line. Some SN candidates are detected in those area.
}
\label{fig:Pointing}
\end{figure*}

The Hyper-Suprime Cam (HSC) \citep{Miyazaki2018,Komiyama2018,Furusawa2018,Kawanomoto2018} on the Subaru Telescope is the only instrument mounted on the prime focus among the large (8-10 m) telescopes. It is unique in its wide field of view (1.77 deg$^2$), with 104 charge-coupled devices (CCDs; 4k$\times$2k pixels) providing a 0.168"/pixel scale. The focal ratio of 1.83 makes the HSC the fastest camera among the large telescopes. An international team, composed of the astronomical communities of Japan, Taiwan and Princeton University,
is in the process of completing a 300-night, 5-year HSC Subaru Strategic Program survey (HSC-SSP, 2014--2019, \cite{aihara18ssp}). Time-domain science is one of the main objectives of this SSP. In this paper, we provide an overview of the HSC transient survey for the COSMOS field.

The HSC transient survey is unique in its depth ($\sim$26 mag) and volume (1.77 deg$^{2}$ FoV), as it explores the deepest transients sky for the area $>1$ deg$^2$ (see Figure \ref{fig:survey} for comparison with other transient surveys).
The HSC-SSP survey has two ultra-deep fields: COSMOS (Cosmic Evolution Survey, \cite{scoville2007cosmos}) and SXDS (Subaru/XMM-Newton Deep Survey, \cite{Furusawa2008}). A cadenced observation will be conducted on these two fields to achieve various transient science cases. In this paper, we will introduce some examples such as Type Ia and Type II-P SN cosmology and SLSNe.

This paper is organized as follows. In Section 2, we describe the survey strategy, data reduction, and transient findings. We show transient samples in Section 3 and present science highlights in Section 4. Finally, we give a summary in Section 5.

\section{Overview of the Transient Survey}
\subsection{Observations}

\begin{figure}
\begin{center}
\includegraphics[width=\columnwidth]{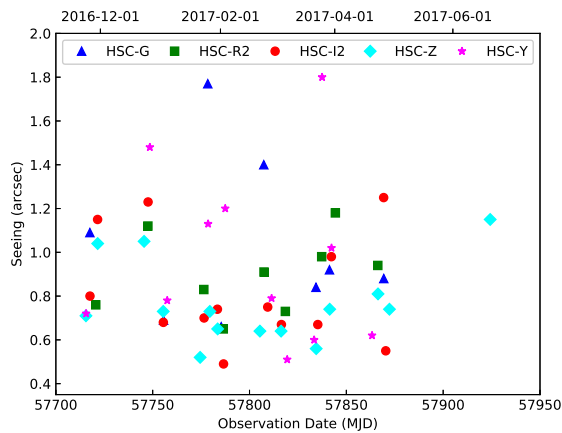}
\end{center}
\caption{Seeing values of each observation.}
\label{fig:Obslog_Seeing}
\end{figure}

Observations were conducted as part of the HSC-SSP \citep{aihara18ssp} from November 2016 to April 2017 on the COSMOS field. 
As shown in Figure \ref{fig:Pointing}, there was one pointing of the ultra-deep layer (solid circle) and four pointings of the deep layer (dashed circles) surrounding the ultra-deep pointing, with significant overlap. The basic observing strategy was to obtain two epochs separated by 7--10 days in all five broad bands ($g$, $r$, $i$, $z$, and $y$-bands) during each monthly observation run for the ultra-deep layer. In total, we acquired 12 epochs over 6 months for each of five bands. For the deep layer, six epochs were obtained over 4 months, with shorter exposure times than the ultra-deep run, as the total planned exposure time was limited. During this transient survey, wide-layer observation around the COSMOS field was also conducted and the data obtained were included in the analysis. Detailed observation dates and typical exposure times are summarized in Table \ref{tab:obslog}. Each image was taken with five-point dithering, as described in \citet{aihara18ssp}, to fill the gaps between the HSC CCDs. There are overlaps between ultra-deep and deep pointings, as mentioned earlier; additionally, the exposure time and other statistics vary from one position to the next. In Table \ref{tab:obslog}, values around the center of the pointings are presented. Seeing values measured on coadded images are listed in Table \ref{tab:obslog} and are shown in Figure \ref{fig:Obslog_Seeing}. 
\begin{longtable}{lll|rrr|rrr}
 \caption{Observation log.} \label{tab:obslog}
 \hline
 &&& \multicolumn{3}{c}{Ultra-deep layer (1.77 deg$^2$)} & \multicolumn{3}{c}{Deep layer (5.78 deg$^2$)} \\
Obs. date  & MJD & filter & exptime & seeing & lim. mag & exptime & seeing & lim. mag \\
 &&& (sec)   & (arcsec) & (mag) & (sec) & (arcsec) & (mag) \\
  \hline
 \endfirsthead
  \hline
  &&& \multicolumn{3}{c}{Ultra-deep} & \multicolumn{3}{c}{Deep} \\
Obs. date  & MJD & filter & exptime & seeing & depth & exptime & seeing & depth \\
 &&& (sec)   & (arcsec) & (mag) & (sec) & (arcsec) & (mag) \\
  \hline
 \endhead
  \hline
 \endfoot
  \hline
 \endlastfoot
\hline
\hline
ref        & 57177.75 & HSC-G  &  7500 & 0.81 &       &  900 & 0.83 & \\
ref        & 57087.75 & HSC-R  &  8460 & 0.63 &       & 1260 & 0.66 & \\
ref        & 57048.44 & HSC-I  & 11730 & 0.60 &       & 2700 & 0.57 & \\
ref        & 57150.73 & HSC-Z  & 28260 & 0.59 &       & 3510 & 0.59 & \\
ref        & 57150.19 & HSC-Y  & 12960 & 0.63 &       & 1890 & 0.62 & \\
\hline
2016-11-23 & 57715.54 & HSC-Z  &  4200 & 0.71 & 25.64  &&&\\
2016-11-23 & 57715.62 & HSC-Y  &  4500 & 0.72 & 25.24  &&&\\
2016-11-25 & 57717.57 & HSC-G  &  1800 & 1.09 & 25.66  &&&\\
2016-11-25 & 57717.62 & HSC-I2 &  3000 & 0.80 & 26.01  &&&\\
2016-11-28 & 57720.60 & HSC-R2 &  1800 & 0.76 & 26.66  &&&\\
2016-11-29 & 57721.55 & HSC-I2 &  2400 & 1.15 & 25.81  &&&\\
2016-11-29 & 57721.60 & HSC-Z  &  3000 & 1.04 & 25.47  &&&\\
2016-12-23 & 57745.56 & HSC-Z  &  5440 & 1.05 & 25.32  & 2070 & 1.04 & 24.83  \\
2016-12-25 & 57747.53 & HSC-R2 &  1560 & 1.12 & 25.88  &  540 & 1.14 & 25.45  \\
2016-12-25 & 57747.62 & HSC-I2 &  3840 & 1.23 & 25.64  &  810 & 1.20 & 25.01  \\
2016-12-26 & 57748.53 & HSC-Y  &  3540 & 1.48 & 22.88 &   810 & 1.39 & 22.36  \\
2017-01-02 & 57755.45 & HSC-Z  &  3300 & 0.73 & 25.59  &&&\\
2017-01-02 & 57755.51 & HSC-I2 &  3000 & 0.68 & 26.51  &  600 & 0.67 & 25.64  \\
2017-01-02 & 57755.61 & HSC-G  &  1710 & 0.69 & 26.75  &  840 & 0.68 & 26.21  \\
2017-01-04 & 57757.52 & HSC-Y  &  4640 & 0.78 & 24.69  & 1610 & 0.69 & 24.27  \\
2017-01-21 & 57774.50 & HSC-Z  &  5240 & 0.52 & 26.31  & 2650 & 0.54 & 25.68  \\
2017-01-23 & 57776.41 & HSC-R2 &  2280 & 0.83 & 26.44  & 1440 & 0.84 & 25.98  \\
2017-01-23 & 57776.54 & HSC-I2 &  3470 & 0.70 & 26.43  & 1510 & 0.70 & 25.84  \\
2017-01-25 & 57778.45 & HSC-G  &  6480 & 1.77 & 26.13  & 2520 & 1.70 & 25.60  \\
2017-01-25 & 57778.62 & HSC-Y  &  3570 & 1.13 & 23.86  &  810 & 1.19 & 23.20  \\
2017-01-26 & 57779.52 & HSC-Z  &   400 & 0.73 & 24.36  &  200 & 0.74 & 24.28  \\
2017-01-30 & 57783.43 & HSC-I2 &  2670 & 0.74 & 26.03  &  810 & 0.75 & 25.31  \\
2017-01-30 & 57783.55 & HSC-Z  &  5250 & 0.65 & 25.85  & 1350 & 0.64 & 25.22  \\
2017-02-01 & 57785.39 & HSC-G  &  3540 & 0.66 & 26.38  & 1440 & 0.66 & 25.77  \\
2017-02-02 & 57786.45 & HSC-R2 &  1380 & 0.65 & 26.60  &  540 & 0.63 & 25.97  \\
2017-02-02 & 57786.59 & HSC-I2 &   800 & 0.49 & 25.83  &  600 & 0.48 & 25.65  \\
2017-02-03 & 57787.47 & HSC-Y  & 10710 & 1.20 & 24.71  & 2430 & 1.25 & 23.96  \\
2017-02-21 & 57805.37 & HSC-Z  &  3840 & 0.64 & 25.69  & 1350 & 0.69 & 25.04  \\
2017-02-23 & 57807.37 & HSC-G  &  3660 & 1.40 & 26.30  & 1440 & 1.31 & 25.86  \\
2017-02-23 & 57807.48 & HSC-R2 &  1680 & 0.91 & 26.33  &  720 & 0.95 & 25.77  \\
2017-02-25 & 57809.40 & HSC-I2 & 10350 & 0.75 & 25.85  & 3240 & 0.64 & 25.21  \\
2017-02-27 & 57811.40 & HSC-Y  &  1800 & 0.79 & 24.44  &  270 & 0.88 & 23.70  \\
2017-03-04 & 57816.31 & HSC-Z  &  3840 & 0.64 & 25.73  & 1080 & 0.68 & 24.81  \\
2017-03-04 & 57816.47 & HSC-I2 &  3210 & 0.67 & 26.33  &  810 & 0.65 & 25.68  \\
2017-03-06 & 57818.51 & HSC-R2 &  1740 & 0.73 & 26.47  &  540 & 0.74 & 25.89  \\
2017-03-07 & 57819.47 & HSC-Y  &  5110 & 0.51 & 25.21  & 1810 & 0.59 & 24.45  \\
2017-03-21 & 57833.38 & HSC-Y  &  4110 & 0.60 & 24.97  &  810 & 0.55 & 24.21  \\
2017-03-22 & 57834.32 & HSC-G  &  2490 & 0.84 & 26.74  & 1215 & 0.80 & 26.22  \\
2017-03-22 & 57834.43 & HSC-Z  &  4380 & 0.56 & 25.82  & 1350 & 0.59 & 25.12  \\
2017-03-23 & 57835.26 & HSC-I2 &  2340 & 0.67 & 25.89  &  540 & 0.65 & 25.30  \\
2017-03-25 & 57837.27 & HSC-R2 &  2220 & 0.98 & 26.11  &  720 & 0.98 & 25.58  \\
2017-03-25 & 57837.49 & HSC-Y  &  1500 & 1.80 & 22.86  &&&\\
2017-03-29 & 57841.29 & HSC-G  &  1740 & 0.92 & 26.53  &  540 & 1.01 & 25.73  \\
2017-03-29 & 57841.41 & HSC-Z  &  3300 & 0.74 & 25.60  &&&\\
2017-03-30 & 57842.27 & HSC-I2 &  3600 & 0.98 & 26.02  &&&\\
2017-03-30 & 57842.34 & HSC-Y  &  4500 & 1.02 & 24.64  &&&\\
2017-04-01 & 57844.33 & HSC-R2 &  3300 & 1.18 & 26.13  &&&\\
2017-04-20 & 57863.28 & HSC-Y  &  4200 & 0.62 & 24.42  &&&\\
2017-04-23 & 57866.25 & HSC-R2 &  2400 & 0.94 & 26.10  &&&\\
2017-04-23 & 57866.36 & HSC-Z  &  3600 & 0.81 & 25.32  &&&\\
2017-04-26 & 57869.27 & HSC-I2 &  4800 & 1.25 & 25.70  &&&\\
2017-04-26 & 57869.33 & HSC-G  &  1200 & 0.88 & 26.45  &&&\\
2017-04-27 & 57870.35 & HSC-I2 &  1800 & 0.55 & 26.09  &&&\\
2017-04-29 & 57872.26 & HSC-Z  &  3300 & 0.74 & 25.30  &&&\\
2017-06-20 & 57924.28 & HSC-Z  &   900 & 1.15 & 23.95  &&&\\
\end{longtable}

\subsection{Data reduction}

HSC pipeline \citep{bosch2018pipeline} version 4.0.5, with default configuration parameters, was used for the data reduction.
We applied the same reduction procedure with the HSC-SSP data release \citep{aihara18dr} for the standard image reduction, which included bias, dark, flat, and fringe corrections, as well as astrometric and photometric calibrations against the PS1 catalog \citep{magnier13}. Typical photometric accuracy is 1\%-2\% \citep{aihara18dr}. Based on the astrometric solutions, the images were warped into a predefined sky grid, which we refer to as a warped image.
Data were processed on a nightly basis. 

For the image subtraction, the difference imaging method of Alard \& Lupton \citep{alard98,alard00} was applied using reference deep, coadded images created from data taken during March 2014 and April 2016. Images with seeing better than 0.7 arcsec were used as reference images, with the exception of the $g$-band. Table \ref{tab:obslog} also includes exposure time and seeing of reference images. Difference imaging was conducted for each warped image, and warped difference images were coadded to create deep difference images for each filter and epoch. With this method, we can avoid subtraction error caused by a discrete change in the point spread function (PSF) at the CCD gaps in coadded images. 

Note that $r$- and $i$-band filters were replaced with filters having a more spatially uniform response across the focal plane in 2016 July and February, respectively \citep{aihara18ssp}. Thus, reference images were observed with the old filters, and search images were observed with the new filters. The non-uniformity in the old filters can result in 4-5\% magnitude offset for high-redshift SNe Ia, whereas the new filters reduced the offset to less than 1\%. A detailed software patch 
is currently under development to match the precision required for SN cosmology. However, the offset described 
did not affect any of the results presented in this paper.

Once coadded difference images are created, we can detect and measure sources in difference images. Based on these sources, transient sources can be identified (see section \ref{sec:transientFinding} for details). Forced photometry was performed at the location of transients for images of all filters and epochs.
Here, the location of a transient was measured by taking the mean of positions in each coadded difference image, in which a transient was detected.

\begin{figure}
  \begin{center}
     \includegraphics[width=\columnwidth]{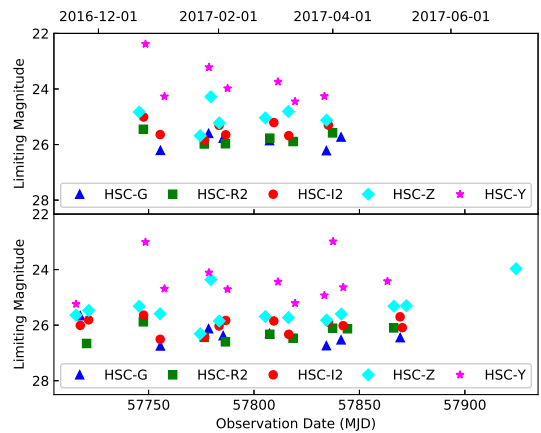}
  \end{center}
  \caption{%
Limiting magnitude (for each filter)
based on 50\% detection efficiency. Upper and 
lower panels correspond to deep and ultra-deep layers, respectively.
 }%
  \label{fig:limitmag}
\end{figure}

\begin{figure}
  \begin{center}
     \includegraphics[width=\columnwidth]{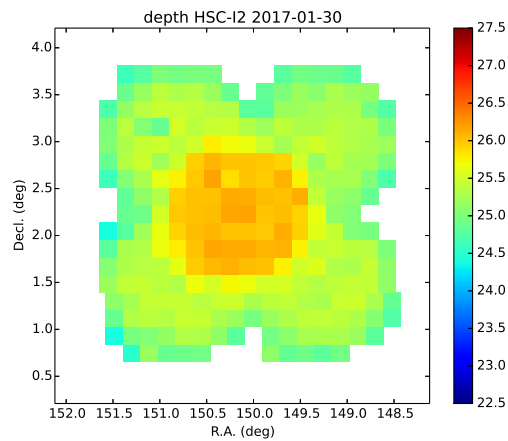}
  \end{center}
  \caption{%
Spatial variation of the detection depth in the $i$-band for a representative night (2017 January 30).
}%
  \label{fig:efficiency}
\end{figure}

\begin{figure}
  \begin{center}
     \includegraphics[width=\columnwidth]{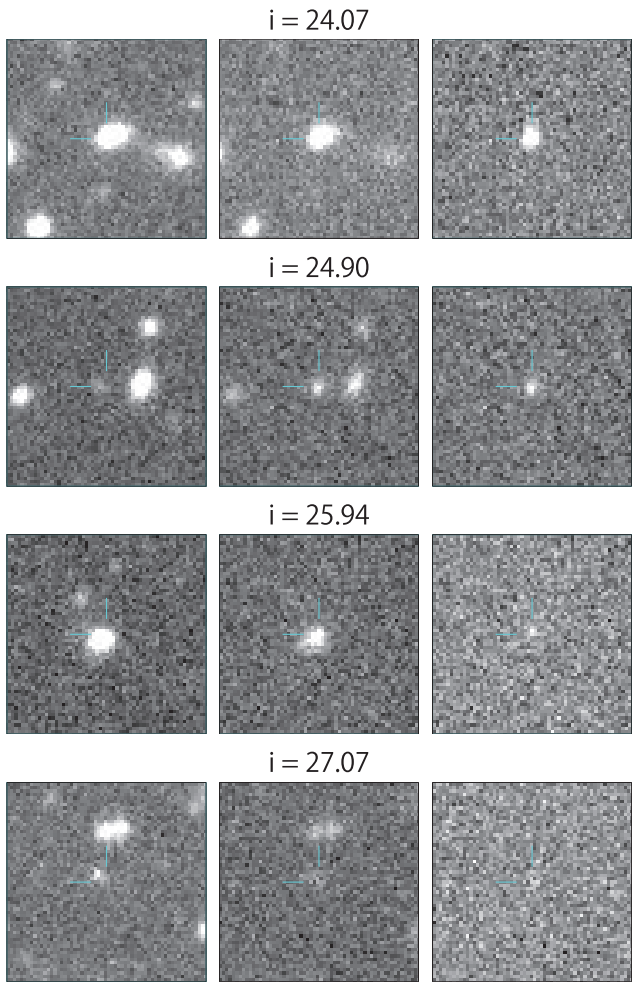}
  \end{center}
  \caption{%
Example of image subtraction of artificial stars in galaxies of the ultra-deep layer. The three images represent a reference image (left), an image with an artificial star in the HSC-I2 band on 2017-01-30 (middle), and a subtracted image (right). The limiting magnitude in the HSC-I2 band on 2017-01-30 is 26.03 (Table~\ref{tab:obslog}).
}%
  \label{fig:fakeongalaxy}
\end{figure}

\subsection{Limiting magnitude/detection efficiency}

To estimate the limiting magnitude of each epoch image, we injected artificial stars with magnitudes ranging from 21 to 28 mag in processed CCD images, before warping to the predefined grid.
The number density of artificial stars was 20,000 / deg$^2$, which corresponds to about 400 objects per CCD. The corresponding CCD images were processed in the same way as the real data. The source catalog from difference coadded images was then compared with the input artificial star catalog. Figure \ref{fig:limitmag} shows the magnitudes at $50\%$ detection efficiency for each filter as a function of the observing epoch. 
Figure \ref{fig:efficiency} shows the spatial distribution of the limiting magnitude of the $i$-band on a specific night. Image subtraction examples for artificial stars embedded onto galaxies are shown in Figure~\ref{fig:fakeongalaxy}; the image subtraction technique worked well for them also.

\begin{figure*}
  \begin{center}
     \includegraphics[width=2\columnwidth]{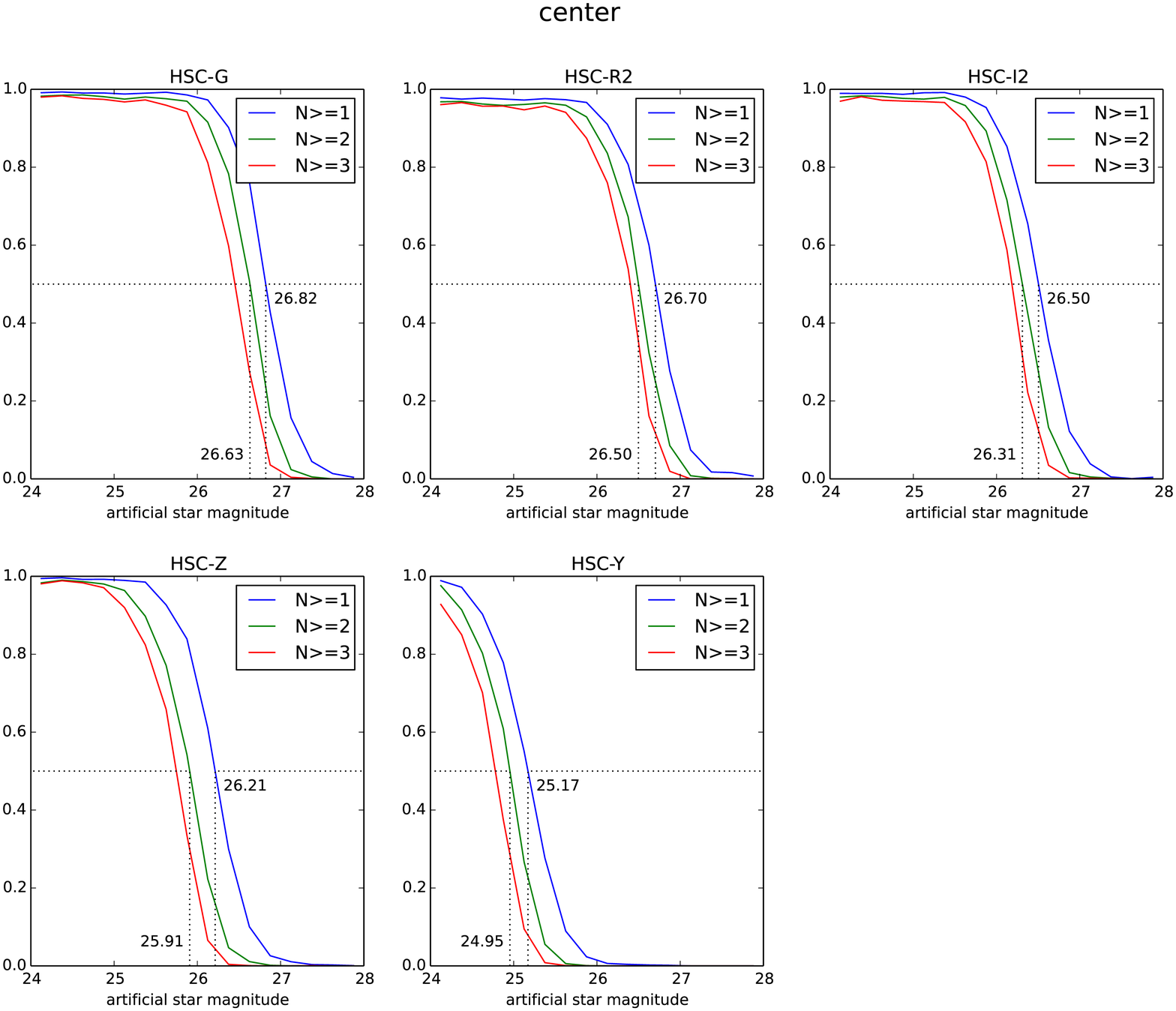}
  \end{center}
  \caption{%
  Detection efficiency as a function of input magnitude. Different curves correspond to a different number of detections.
}%
  \label{fig:detectionrate}
\end{figure*}

For transient findings, we imposed at least two detections at the same position (see section \ref{sec:transientFinding}). Figure \ref{fig:detectionrate} shows the detection rate of constant brightness objects as a function of magnitude. Different lines denote how many times the same object was detected at the same position.

\begin{figure}
  \begin{center}
     \includegraphics[width=\columnwidth]{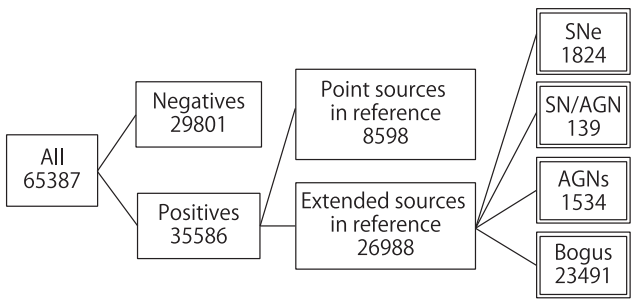}
  \end{center}
  \caption{%
Flow chart of transient classification. "All" includes the transients after the real/bogus judgment by the convolutional neural network.}
  \label{fig:flowchart}
\end{figure}

\begin{table*}
  \tbl{Number of candidates classified as SN, SN/AGN, AGN, and bogus after screening with AUC boosting and partial AUC optimization.}{%
  \begin{tabular}{lcccc}
  	\hline
    Class & CNN & CNN+AUCB & CNN+PAUC & CNN+AUCB+PAUC\\ 
     & & $N$(CNN+AUCB)/$N$(CNN) & $N$(CNN+PAUC)/$N$(CNN) & $N$(CNN+AUCB+PAUC)/$N$(CNN)\\ \hline
    SN & 1824 & 1728 & 1724 & 1666\\
    &&94.7\%&94.5\% & 91.3\%\\
    SN/AGN & 139 & 129& 134 & 128\\
    &&92.8\%&96.4\%&92.0\%\\
    AGN & 1534 & 1472 & 1492 & 1450\\
    &&95.9\%&97.2\%&94.5\%\\
    Bogus & 23491 & 13149 & 16518 & 11761\\
    &&56.0\%&70.3\%&50.1\%\\
    \hline
  \end{tabular}}\label{tab:ml}
\end{table*}

\subsection{Transient finding}
\label{sec:transientFinding}
Sources detected on coadded difference images have been classified as real or bogus by machine learning techniques. We adopted a convolutional neural network (CNN) with a combination of convolution, pooling, and dropout layers, trained by 100,000 artificial stars as a real sample and 100,000 objects within actual observational images as a bogus sample. The trained CNN was validated with 10,000 artificial stars and 10,000 bogus samples. Note that here, a bogus sample includes ``real'' transients, as they are taken from actual observational data.
The CNN showed a false-positive rate of 4.3\% and 6.0\% at the true-positive rate of 90\% for objects with a signal-to-noise ratio better than 10 and 7.5, respectively. These values are not necessarily better than those cited in a previous study for the HSC data \citep{morii16} using various machine learning methods for measured parameters. This is because our bogus sample was constructed from actual observational data and included a large number of "real"  transients, as the reference images are taken well before the search observation. The actual performance is expected to be better than that indicated by these values. We also applied the area under the curve (AUC) boosting method \citep{morii16} and partial AUC optimization \citep{Ueda2018} for real/bogus classification. The results are discussed in section \ref{sec:classification}.

After the CNN screening, if the same source was identified at the same place (within $0.4$ arcsec), then that source was registered as a transient. In total, 65,387 transient candidates were identified.
For the registered transients, the closest object in reference images was tagged as the host object. Sometimes, the host object identification is not optimal (matched with very faint noise-like objects or matched with deblended children of big galaxies); in such cases, any clear misidentifications were corrected by visual inspection. The host objects were then matched with a compilation of public redshift catalogs, COSMOS2015 \citep{laigle2016cosmoscat}, and HSC photo-z catalog \citep{tanaka18photz} objects with a search radius of $0.5$ arcsec. Public redshift catalogs included SDSS DR12 \citep{SDSS_DR12}, PRIMUS DR1 \citep{PRIMUS_DR1_1,PRIMUS_DR1_2}, VVDS \citep{VVDS}, zCOSMOS DR3 \citep{zCOSMOS_DR3}, and 3D-HST \citep{3DHST_1,3DHST_2}.


\begin{figure*}
  \begin{center}
     \includegraphics[width=2\columnwidth]{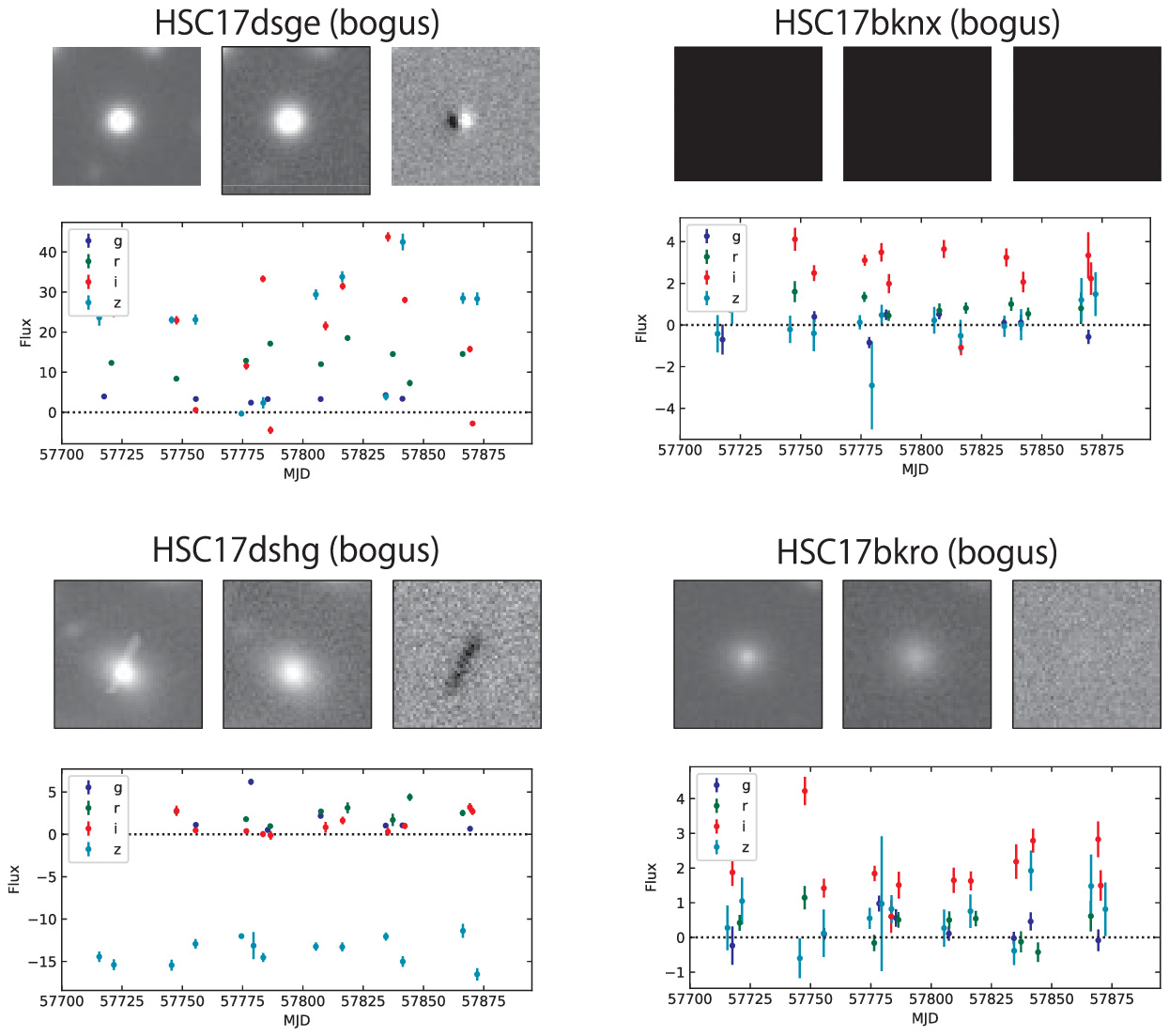}
  \end{center}
  \caption{%
Examples of images and light curves of bogus objects. The three images are a reference image (left), new image (middle), and subtracted image (right). HSC17dsge is a bogus-object example caused by imperfect subtraction around a bright star, which is not classified as a "Point source in reference" due to the relatively large distance. HSC17dshg has an artifact in the $z$-band reference image. HSC17bknx shows a mis-subtraction near the center of a galaxy in the $i$-band image. HSC17bkro is an example of an object that was detected only twice, with a low signal-to-noise ratio.}%
  \label{fig:bogusfigure}
\end{figure*}

\begin{figure*}
  \begin{center}
     \includegraphics[width=2\columnwidth]{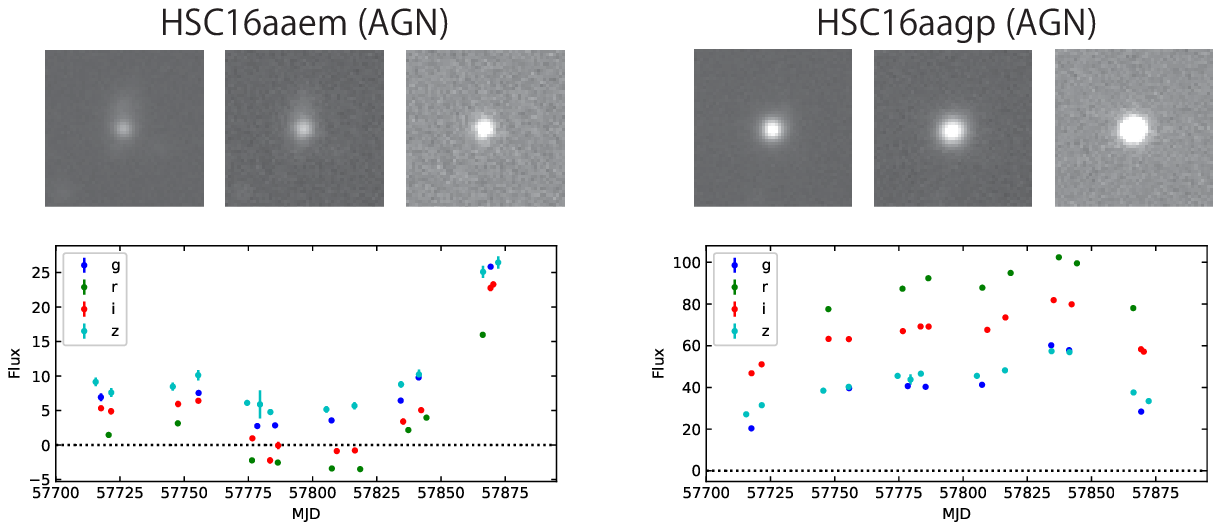}
  \end{center}
  \caption{%
Examples of images and light curves for active galactic nuclei (AGN).}   
\label{fig:agn}
\end{figure*}

\section{Sample of Transients}
\subsection{Classification of transients}
\label{sec:classification}
We mainly focus on extragalactic transients and SNe in this paper. To identify SN candidates, 65,387 candidates were classified according to their properties after CNN screening. The flowchart of the classification is shown in Figure~\ref{fig:flowchart}. First, we excluded objects having light curves dominated by negative PSF fluxes. When a light curve of a certain filter had more negative points than positive points, it was judged to be a negative light curve. If the number of filters with a negative light curve exceeded that with a positive light curve, the object was excluded as a ``Negative" candidate. Second, we checked host objects in reference images. If the host object was a point source, we excluded it, as it was most likely to be a variable star. 
Note that some active galactic nuclei (AGN) are classified as ``Negative" and ``Point source in reference".
Finally, visual inspection was performed on the remaining 26,988 candidates by nine experts.
The light curve shapes were also visually checked, along with the time series of coadded difference images.

In the visual inspection, we first excluded any bogus detections, which accounted for a large proportion of the remaining candidates (Figure \ref{fig:bogusfigure}). 
Some of the bogus detections are caused by imperfect image subtraction near bright stars (see HSC17dsge in Figure \ref{fig:bogusfigure}); this occurs because they are not classified as a "Point source in reference" if the offset from the center of the star is relatively large. A large portion of these objects was detected in only one or two bands, with nearly constant fluxes (see HSC17bknx); this may occur due to imperfect subtraction. Other bogus detection examples include artifacts in the reference image (HSC17dshg) and too few detections with low signal-to-noise ratios (HSC17brko).

\begin{figure}
  \begin{center}
       \includegraphics[width=\columnwidth]{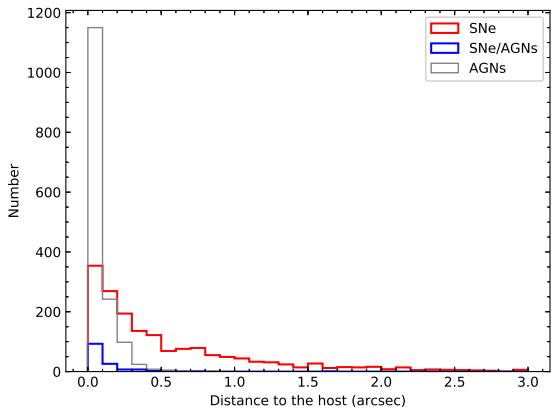}
  \end{center}
  \caption{%
Distributions of the distance to the host object in reference images. Red, purple, and gray lines show the SN, SN/AGN, and AGN distributions, respectively.
}%
  \label{fig:radial_distribution}
\end{figure}

We then classified the clean objects into SN or AGN. When a candidate had a clear offset ($>3\sigma$ of centroid error) with a host object, it was classified as a SN. AGN can be identified by their association with an X-ray source using the COSMOS2015 catalog. However, because some SNe can occur in X-ray bright galaxies and low-luminosity AGN can elude X-ray detection, we mainly classified AGN based on the light curve shapes (e.g., multiple peaks or a very long duration; see Figure \ref{fig:agn}).
Ultimately, 1,824 objects were classified as SN, and 1,534 as AGN. Marginal cases (139 objects) were flagged as ``SN/AGN". Figure \ref{fig:radial_distribution} shows the distributions of the apparent distance to the host object. AGN were almost exclusively located within $\sim 0.2$ arcsec from the host objects. On the other hand, SNe were distributed more widely. 
Note that our AGN samples are not complete. Complete AGN samples including candidates classified as ``Negative" and ``Point source in reference" will be presented in forthcoming papers.

Although visual inspection was performed on all of the objects following CNN screening and the exclusion of "Negative" and "Point source in reference" objects, we also applied the real/bogus classification process with AUC boosting \citep{morii16} and partial AUC optimization, using a deep neural network \citep{Ueda2018}.
In Table \ref{tab:ml}, we show the results only for classified objects to compare the reduction factors of SNe, AGN, and bogus objects. By applying AUC boosting and partial AUC optimization, in addition to CNN, the number of bogus objects was reduced to 56\% and 70\%, while more than 94\% and 94\% of real objects (SNe and AGN) were preserved, respectively. When we applied both methods, the bogus objects were reduced to 50\%, with more than 91 \% of the real objects retained. This highlights the usefulness of multiple real/bogus classifiers, as demonstrated in \citet{morii16}.

\begin{figure}
  \begin{center}
       \includegraphics[width=\columnwidth]{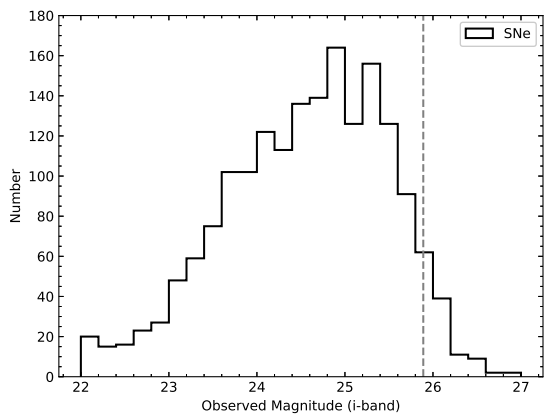}
  \end{center}
  \caption{%
Distribution of the $i$-band peak magnitudes for our SN samples.
Dashed line shows the median depth in the $i$-band.
}%
  \label{fig:mobs_distribution}
\end{figure}

\begin{figure}
  \begin{center}
       \includegraphics[width=\columnwidth]{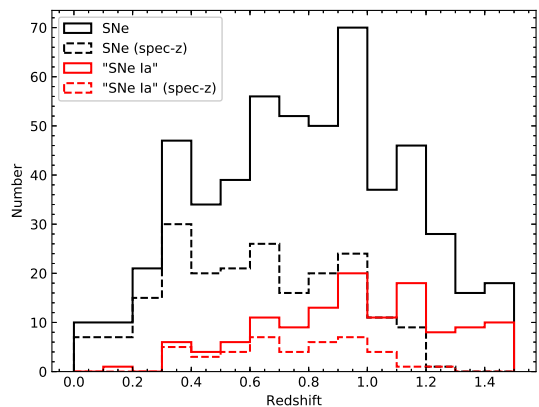}
  \end{center}
  \caption{%
Distribution of redshifts for our SN samples
with spec-z (207 objects) or COSMOS photo-z (381 objects) of the host objects (black solid).
The distribution for objects classified as ``SNe Ia'' by SALT2 fitting using spec-z or COSMOS photo-z (129 objects) is shown with a red solid line.
Dashed lines show the spec-z samples.
}%
  \label{fig:redshift_distribution}
\end{figure}

\begin{figure*}
  \begin{center}
  \includegraphics[width=2\columnwidth]{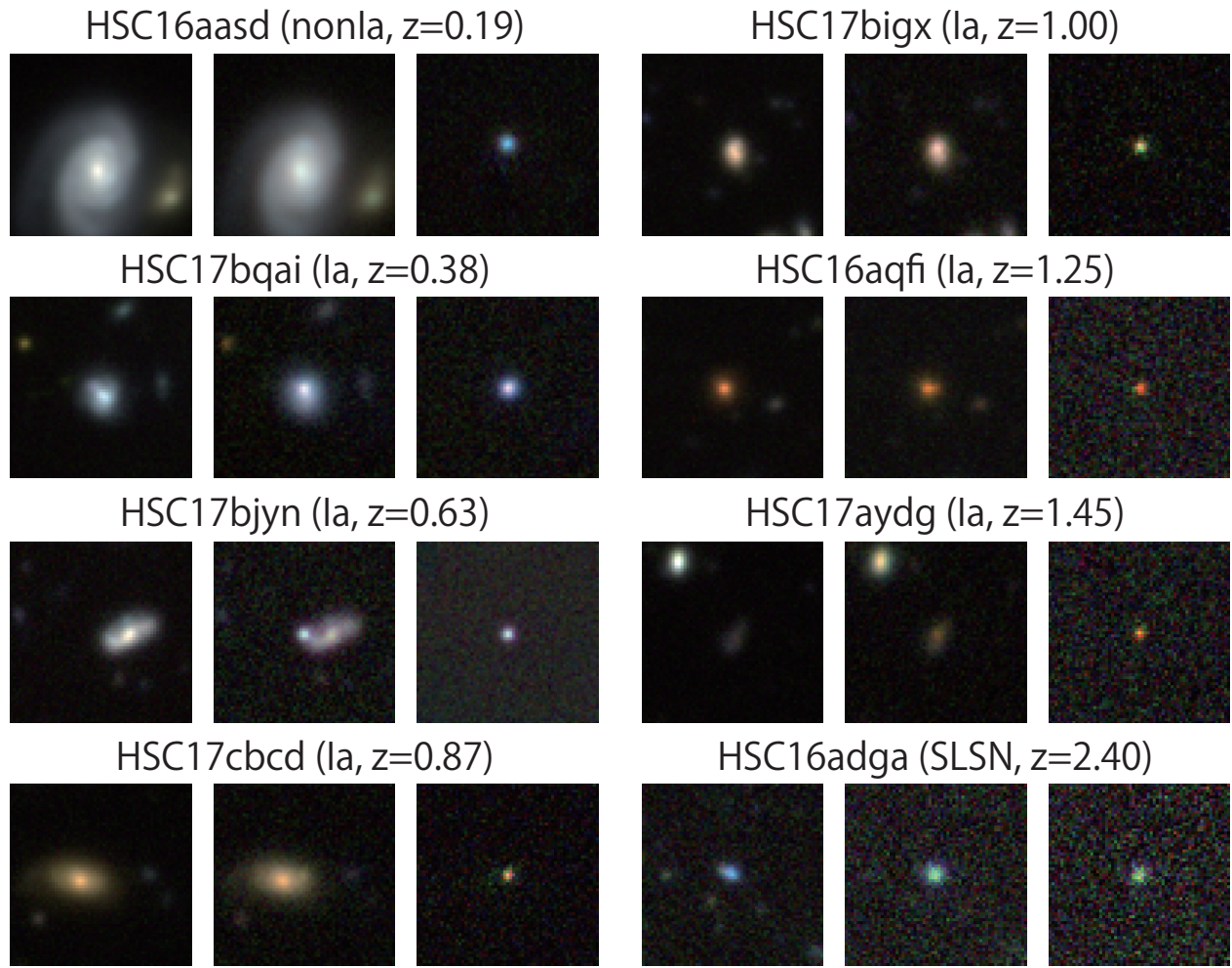}
  \end{center}
  \caption{%
Images of SN candidates at various redshifts. Redshifts are spec-z, except for HSC17aydg (HSC photo-z) and HSC16adga (COSMOS photo-z). Three panels are shown for an SN: reference (left), new image (middle), and subtracted image (right). Three filter-bands (r2-, i2-, and z-bands) make up this color composite.
}%
  \label{fig:image}
\end{figure*}

\subsection{Supernova candidates}

In this section, we examine the properties of objects classified as SN. Figure \ref{fig:mobs_distribution} shows the distribution of observed $i$-band magnitudes at the peak. Here ``peak" refers to the brightest magnitude within the observation and not the result of light curve fitting. For the ``SNe Ia" sample described below, the difference between this ``peak" magnitude and the peak magnitude obtained from the light curve fit shows a $0.08$-mag offset (the "peak" magnitude is fainter) and $0.15$-mag scatter. A majority of our samples exhibit $24.0$--$25.5$ mag at the peak, with a tail to $26$--$27$ mag. Thus, the HSC-SSP transient survey is among the deepest transient surveys (see also Figure \ref{fig:survey}), detecting a larger number of SNe than SN surveys with the HST \citep{Dawson2009,rodney14,graur14}.

With deep depth, these transients are located at high redshifts. By virtue of the rich dataset in the COSMOS field, 207 and 371 SN candidates have spectroscopic redshifts (hereafter, spec-z) and COSMOS2015 photometric redshifts (COSMOS photo-z), respectively, by identifying potential host galaxies of the SN candidates. Figure \ref{fig:redshift_distribution} shows the redshift distribution of these 578 objects. The distribution has a median of $z = 0.85$; 187 objects (32\%) are located at $z > 1$. Images of example SNe in each redshift range are displayed in Figure \ref{fig:image}. For 141 of the SN candidates, we were unable to identify a clear host galaxy. A detailed analysis of hostless samples will be described in a separate paper.

For classification of SNe Ia and other types of SNe, we applied the SALT2 light curve fitter \citep{guy2007} to our SN sample. 
SALT2 is an empirical model of SN Ia's spectro-photometric evolution over time. The model is constructed using a large data set that includes light curves and spectra of both nearby and distant SNe, up to redshift 1. Available ultraviolet (UV) spectra from the International Ultraviolet Explorer (IUE) are also included. The model is valid in the spectral range of $2,500-8,000$\AA. The public version of the SALT2 package\footnote{\tt http://supernovae.in2p3.fr/salt/} provides light curve fitting algorithms based on this model. Light curve fitters can estimate $T_{max}$ (time of peak brightness in the $B$-band), redshift, $c$ (the SN color parameter of the model), $x_0$ (the flux scale or luminosity distance), and $x_1$ (the light curve shape parameter of the model) to fit the observed multi-band photometric data by minimizing $\chi^2$. A commonly used fitting tool, {\tt snfit}, fixes the redshift to a given value, allowing the other four parameters to be determined. Another fitter, {\tt snphotoz}, fits all five parameters. At the first stage of {\tt snphotoz}, both $c$ and $x_1$ are fixed to 0 (mean values of SN Ia), and then the redshift is scanned to find the minimum $\chi^2$ value. This redshift is then used as an initial value for a full five-parameter fit. One may question the validity of using the SALT2 model at a redshift beyond 1. \citet{Balland2018} found no evidence of redshift evolution from Very Large Telescope (VLT) spectra of SNe Ia below a redshift of 1.0. There has been no observational study clearly showing the evolution of SN Ia beyond redshift 1, mainly because the observations are limited by the faintness of the objects. In this paper, we simply assumed that the SALT2 model is valid beyond redshift 1.

For all SNe, we ran {\tt snfit} by constraining the redshift to the best available redshift. For SNe that do not associate with a clear host galaxy (hostless), or which have only photometric redshifts from HSC broad-band photometry, we also ran {\tt snphotoz} to estimate both the redshift and light curve parameters. For both {\tt snfit} and {\tt snphotoz}, we added an option ``{\tt -w 2500 8000}'' to use a wider wavelength range. When the reduced $\chi^2$ of {\tt snphotoz} was less than 70\% of the reduced $\chi^2$ of {\tt snfit}, we adopted the result of {\tt snphotoz}. 
Note that the $y$-band images suffer from scattered light, and some of the objects are affected by imperfect correction \citep{aihara18dr}. Therefore, we have not used $y$-band data for light curve fits, to ensure a similar classification process for all SN candidates.

\begin{figure}
  \begin{center}
  \includegraphics[width=\columnwidth]{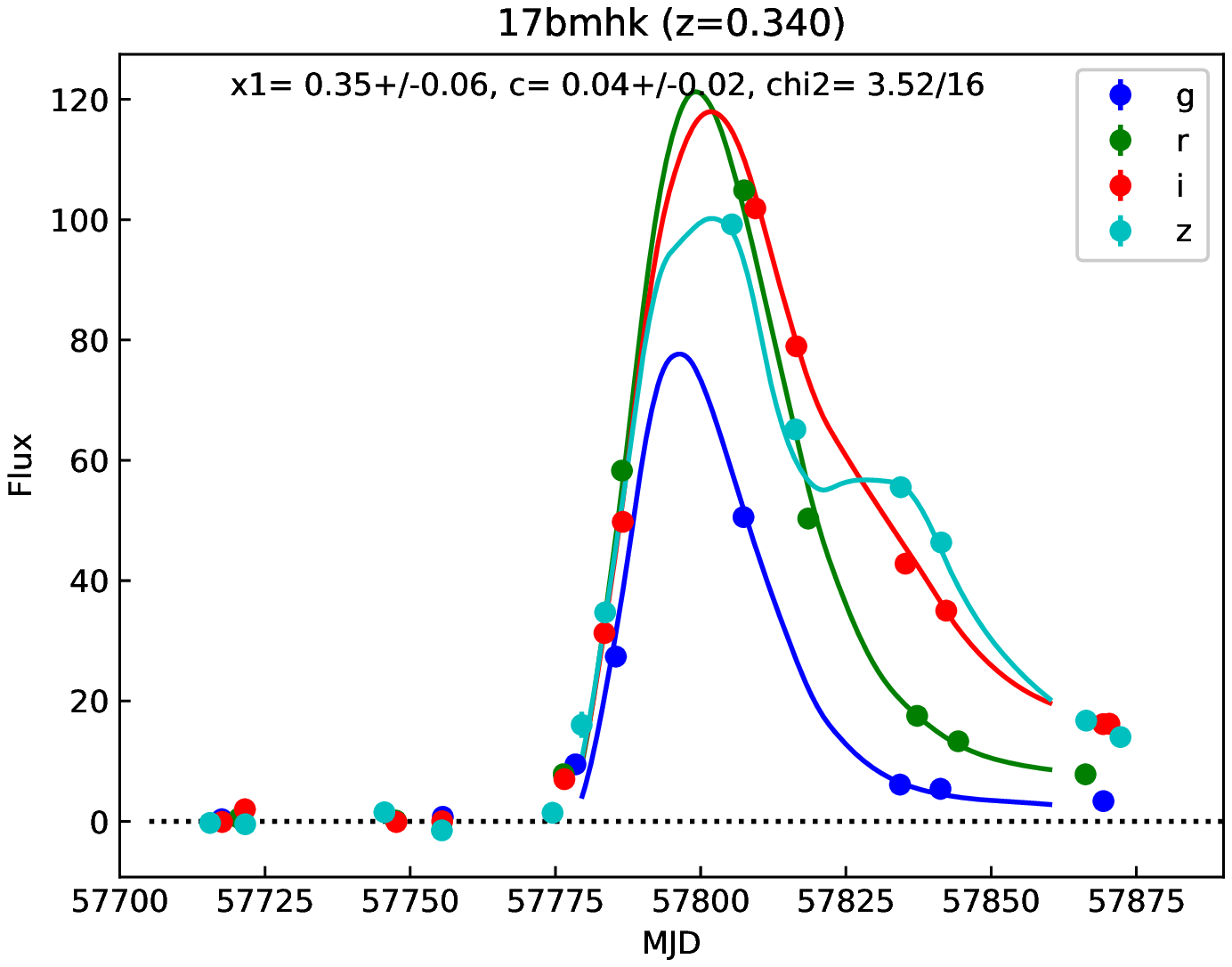}
  \includegraphics[width=\columnwidth]{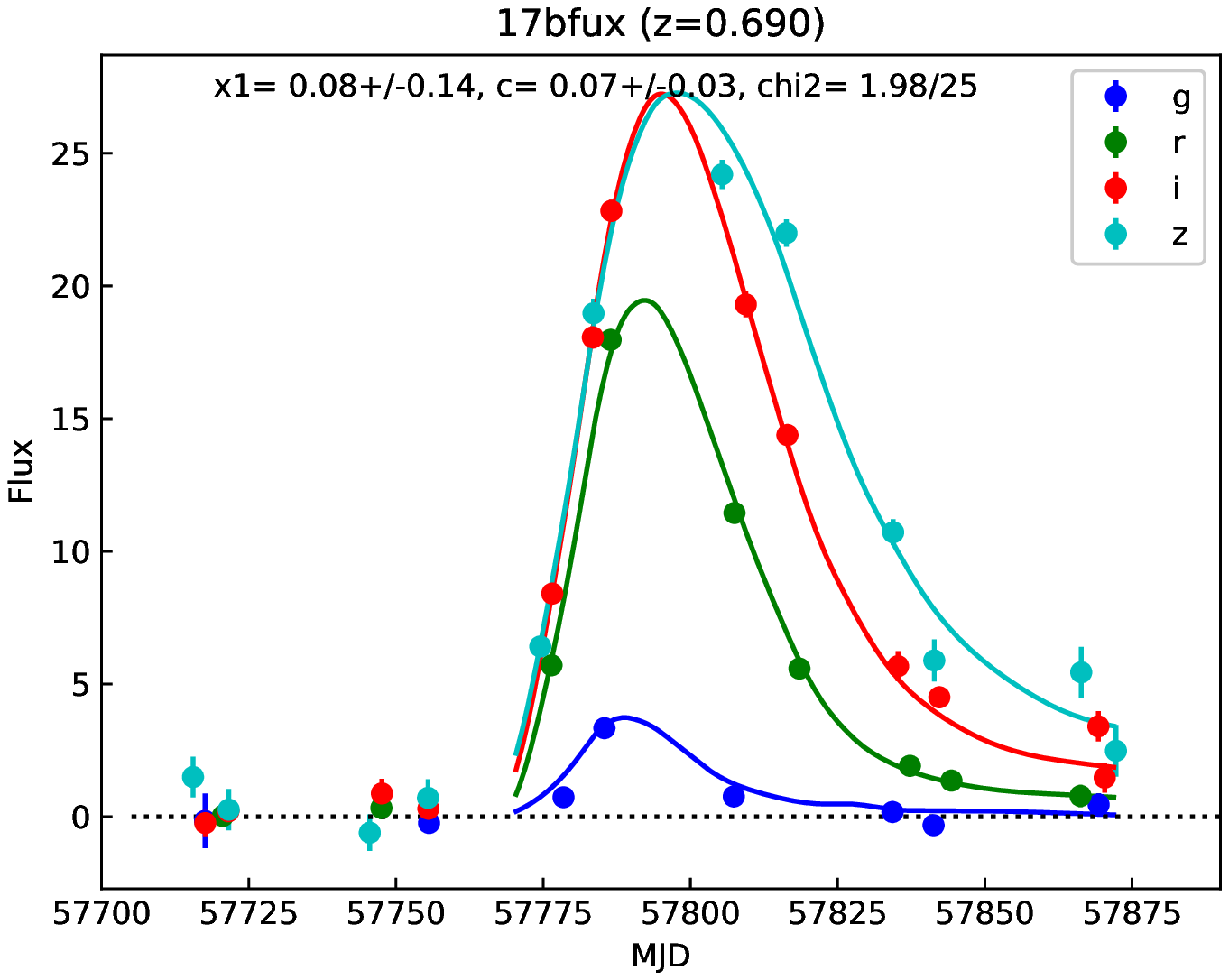}
  \includegraphics[width=\columnwidth]{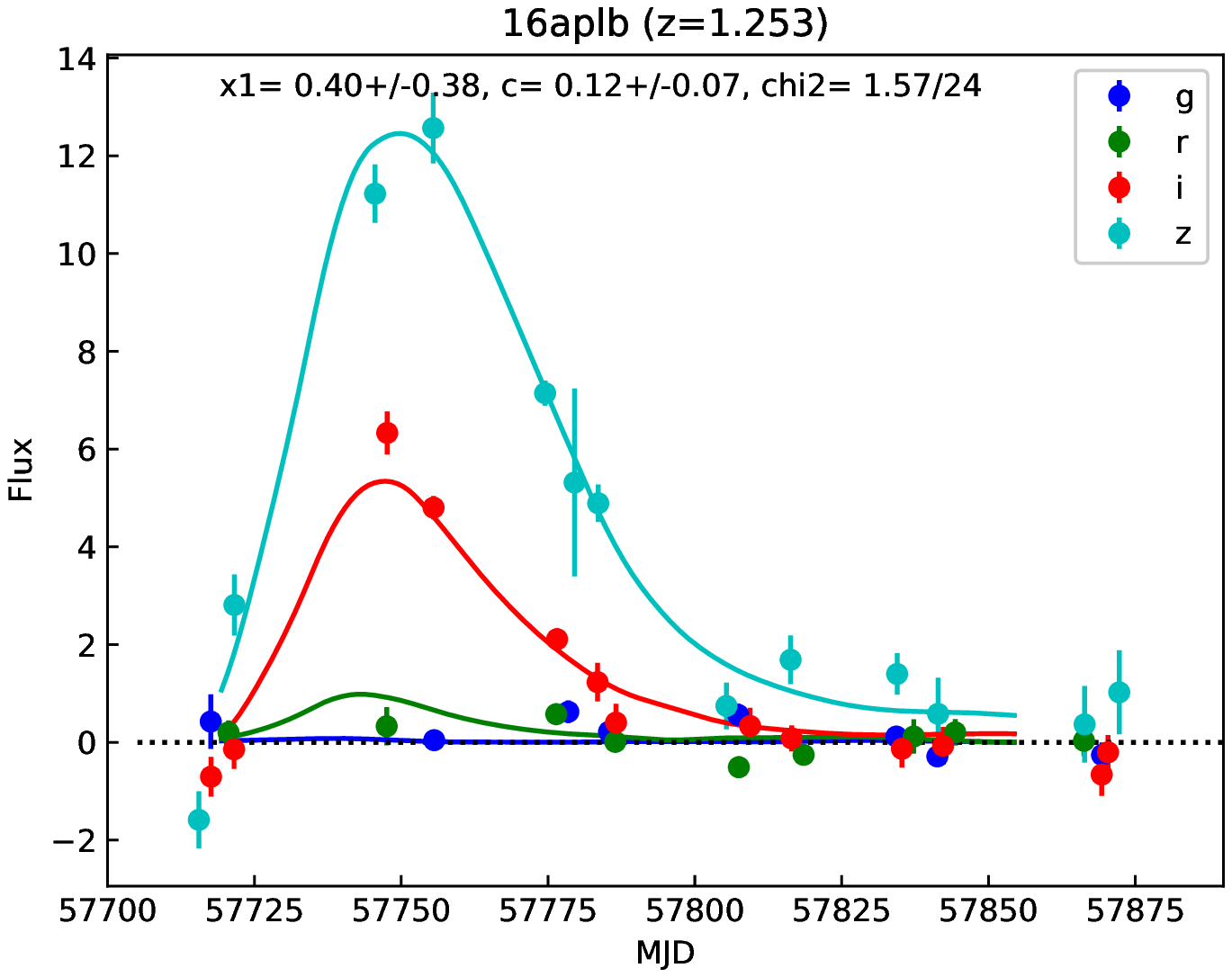}
  \end{center}
  \caption{%
Light curves of three ``SNe Ia" at redshifts of 0.340 (spec-z), 0.690 (spec-z), and 1.253 (COSMOS photo-z).
The ordinate axis represents the point spread function-fitted flux measured on coadded difference images scaled for a zero-point of 27.0 mag.
}%
  \label{fig:SNIa}
\end{figure}

With regard to defining ``SNe Ia" samples in this paper, we selected SNe with the following characteristics: (1) light curve parameters, color ($c$), and shape ($x_1$) within the $3\sigma$ range of \citet{scolnickessler2016}'s ``All G10" distribution, (2) a $M_B$ brighter than $-18.5$ mag, (3) a reduced $\chi^2$ of less than 10, and (4) the number of degrees of freedom is greater than or equal to 5. In total, 433 SNe were classified as ``SNe Ia". Among them, 129 SNe have spec-z or COSMOS photo-z; the relatively small fraction of these with respect to the total number of SN is mainly due to the large area outside of the original COSMOS field, as shown in Figure \ref{fig:Pointing}. For 57 of the SNe with either incorrect or unavailable redshifts, the redshift was recovered by {\tt snphotoz}. Figure \ref{fig:SNIa} shows representative light curves of ``SNe Ia" at different redshifts.

As described above, our ``SN Ia" sample was selected solely from photometric information; no SNe spectroscopic information was used. Thus, this sample may include contamination from other SNe types or may be missing genuine SNe Ia. In this paper, we do not focus on detailed SN classification, as more detailed SN classification will be presented in other papers. The number of "SN Ia" (433) is relatively small compared with the entire sample number (1,824). This is due to our conservative criteria. Infact, there are 232 SNe with a small number of available data points, which do not satisfy condition (4) above. If we loosen condition (1), corresponding to the light curve parameters condition of being within $3\sigma$ to being within $5\sigma$, as well as condition (3), in which the $\chi^2$ condition is reduced from less than 10 to less than 20, 104 additional SNe can be assigned to ``SN Ia". This results in $(433 + 104)/(1824 - 232) = 34\%$ as the SNe Ia fraction. Note that this value is still lower than the general expectation of about $50\%$. The detailed classification algorithms differ; photometric identifications of SNe Ia in SDSS \citep{Sako2011} and Pan-STARRS1 \citep{Jones2017} SN surveys show a similarly low SN Ia fraction. 

\begin{figure}
  \begin{center}
         \includegraphics[width=\columnwidth]{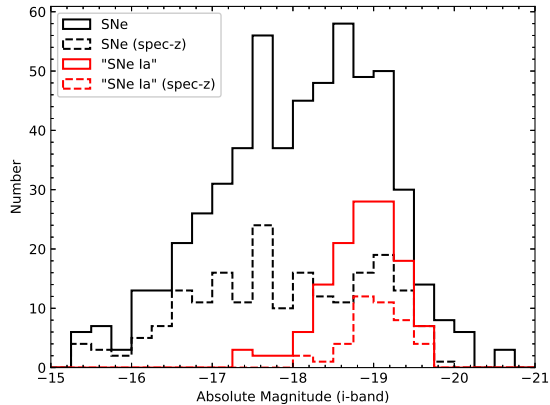}
  \end{center}
  \caption{%
Distributions of absolute magnitudes of our 571 SN samples with spec-z and COSMOS photo-z (black solid), with 129 ``SNe Ia'' among them (red solid). Dashed lines show the samples with spec-z.
}%
  \label{fig:mabs_distribution}
\end{figure}

\begin{figure}
  \begin{center}
         \includegraphics[width=\columnwidth]{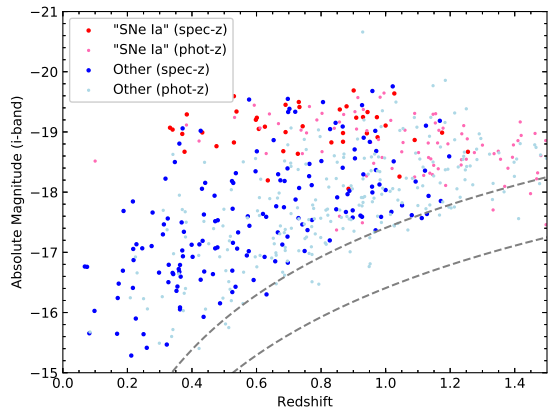}
  \end{center}
  \caption{%
Relation between peak $i$-band absolute magnitude and redshift. Red and blue dots show ``SNe Ia'' and remaining SNe, respectively. Dots with dense and pale colors show objects with spec-z and COSMOS photo-z, respectively. Two dashed curves show constant apparent $i$-band magnitudes of 26 (upper) and 27 (lower) mag.
}%
  \label{fig:mabs_redshift}
\end{figure}

Distributions of absolute magnitudes are shown in Figure \ref{fig:mabs_distribution}; absolute magnitudes as a function of redshift are shown in Figure \ref{fig:mabs_redshift}. In these figures, we only show SNe with spec-z and COSMOS photo-z. The absolute magnitudes of ``SNe Ia'' are clustered at $-18$ to $-19$ mag. The redshift distribution of ``SNe Ia'' has a median of $z = 0.97$, and 58 objects (45\%) are located at $z > 1$. 

Note that some SNe, which are not classified as ``SNe Ia'', have absolute magnitudes brighter than $-19$ mag. Although some of these objects are real (such as bright Type IIn SNe or SLSNe, see Section \ref{sec:highlight}), it should be noted that most of non-``SNe Ia'' at $z > 1$ have only photometric redshifts (Figure \ref{fig:mabs_redshift}), and the magnitude distributions are affected by the uncertainties of the photometric redshifts. In fact, if we limit ourselves to using only spectroscopic redshifts, the absolute magnitudes of non-``SNe Ia'' have a steep cutoff around $-19$ mag; there is no object with an absolute magnitude of $<- 20$ mag.

\section{Science highlights}
We discuss some of the latest findings obtained using the HSC transient survey data presented in this paper.

\label{sec:highlight}

\subsection{Type Ia SN cosmology}
Two decades have passed since the discovery of dark energy \citep{perlmutter99a,riess98a}, yet its nature still remains as one of the biggest mysteries in modern physics. Today, baryon acoustic oscillation  augments the measurement of dark energy, but SN~Ia still leads with respect to measurement precision. Now we would like to know whether dark energy changes in time or not. The HSC transient survey is designed to probe redshifts in $1.0 < z < 1.4$ where time variability becomes sensitive. 

We identified 433 ``SNe~Ia'', based on the light curve and photometric redshifts from host galaxies. In particular, we found 58 ``SNe Ia'' beyond redshift $z > 1.0$, with reliable spec-z or COSMOS photo-z. In the past, only the HST could reach this redshift range, and only two dozen SN~Ia have been measured \citep{suzuki12a,riess18a}. HSC is the only instrument that can probe $z > 1.0$ SNe from the ground having the necessary photometric accuracy for cosmological analysis. By doubling the number of high-redshift SNe~Ia, we expect to impose a tight constraint on the nature of dark energy, in which the cosmological parameter becomes sensitive with high redshifts. 

We conducted a spectroscopic follow-up campaign for live SNe with large telescopes: Subaru/Faint Object Camera and Spectrograph (FOCAS), Keck/Low Resolution Imaging Spectrometer (LRIS), VLT/visual and near-UV FOcal Reducer and low dispersion Spectrograph (FORS), Gemini/Gemini Multi-Object Spectrographs (GMOS), and Grand Telescopio CANARIAS (GTC)/Optical System for Imaging and low Intermediate Resolution Spectroscopy (OSIRIS). 
During the 2017 season, in collaboration with the COSMOS Lyman-Alpha Mapping And Tomography Observations (CLAMATO) project team \citep{lee14a}, we placed a few slits on live SNe on the Keck/LRIS mask, while other slits were used for Lyman break galaxies. We observed 17 live SN spectra; the details will be reported in a forthcoming paper.

We also performed host galaxy spectroscopic follow-up observations. First, we used the COSMOS2015 catalog to determine if spec-z was known; if so, we adopted that redshift \citep{laigle2016cosmoscat}. When slit mask observation was conducted, we placed slits on potential host galaxies in the field of view for Keck/LRIS, Subaru/FOCAS, and VLT/FORS. The most efficient observation was conducted by the 3.9-m Anglo-Australian Telescope (AAT)/AAOmega spectrograph, which has 400 fibers in the two-square-degree field of view. In the 2018 February run, we collected 257 host galaxy spectra, with the goal of completing the collection in the upcoming semesters.

Although the HSC can detect SN~Ia beyond $z > 1.2$, the peak flux of SN~Ia goes into the infrared (IR), and HSC loses its sensitivity. We conducted an IR imaging follow-up observation via the HST.   
For cosmological analysis, we introduced light curve width (stretch) correction and color correction \citep{tripp98b}, and it is essential to reduce error propagation from colors. Together with HSC data, we designed our HST IR follow-up observation to reduce color-associated error to less than 3\%.
High-z SN~Ia candidates were identified from HSC observations. An observing request was sent to the HST. We successfully observed 26 SNe~Ia candidates ($z > 1$) with the HSC and HST, but have yet to observe the HST reference images. We hope to collect spectroscopic redshifts of the host galaxies for cosmological analyses in the near future.

\subsection{Type II SNe}
Type II SNe (SNe II) constitute the most common class among core-collapse SNe \citep{2011MNRAS.412.1441L}, tracing the most typical evolutionary path of massive stars. 
Characterized by hydrogen features, they are robustly interpreted as an explosion of a red supergiant. Analysis of their light curve properties, which thus far have been limited mostly to the local sample, provides a clue as to the nature of the progenitor evolution and actively debated explosion mechanism \citep{2014ApJ...786...67A}. In addition, there has been increasing interest in using SNe II for cosmological applications \citep{2002ApJ...566L..63H,2006ApJ...645..841N,2010ApJ...715..833O,2017ApJ...835..166D}

The SNe light curves are characterized by a rapid increase to the peak value ($\sim 7$ days), followed by a slow decline, frequently showing a plateau, for $\sim 100$ days \citep{2015MNRAS.451.2212G}. As such, the COSMOS HSC transient survey is suited to SNe II discovery and characterization of the entire evolution of their multi-color light curves. The peak in the redshift distribution is expected to occur at $z \sim 0.2$, with extension to $z \sim 0.4$, far beyond the well-established available SN II samples. 

In \citet{dejaeger2017}, we used one SN II discovered by the COSMOS HSC transient survey to extend the SN II Hubble diagram to $z = 0.340$. We applied the so-called standard candle method, which uses a correlation between the intrinsic luminosity and the expansion velocity from the spectra \citep{2002ApJ...566L..63H,2006ApJ...645..841N}. Followed by the rapid rise and a plateau-like evolution of SN 2016jhj, as our best high-redshift SN II candidate with the COSMOS transient survey, spectroscopy was performed with Keck/LRIS; the LRIS spectrum confirmed it to be an SN II.
Cross correlation to SN II template spectra was performed to measure the expansion velocity, in an attempt to place this object within the luminosity-- expansion velocity relationship \citep{2009ApJ...694.1067P}. The standardized magnitude, as well as the color correction (due to extinction), were then modeled with the cosmological parameters using a Monte Carlo Markov Chain simulation, to derive the best-fit Hubble diagram and the probability distribution of the parameters. The derived cosmological parameters are consistent with the $\Lambda$-CDM model. The resulting dispersion in the standardized magnitude was 0.27 mag (i.e., 12-13\% in distance). This work represents a proof-of-concept of the capability of high-redshift SN II cosmology ($z > 0.3$). 

We plan to extend the analysis to the photometric color method \citep{2015ApJ...815..121D,2017ApJ...835..166D}, which uses only photometric information to derive the distances to SNe II. Minimal requirements of this method include light curve information from two bands, the slope of the plateau in a given band pass, and a color term. This methodology fits well to the HSC sample, which consists of a number of (apparently) faint SNe II for which the spectroscopic follow-up is difficult to coordinate. Given the quality of the multi-band light curves in the HSC sample, both in terms of photometric accuracy and coverage over the entire survey duration, the application of this method is straightforward. This approach will make the best use of the large sampling of HSC-discovered high-redshift SNe II, to derive the Hubble diagram up to $z \sim 0.4$.

\subsection{Superluminous SNe}
We searched for high-redshift superluminous SNe (SLSNe) in this survey. High-redshift SLSN candidates were selected based on the photometric redshifts of host galaxies. We mainly used the photometric redshifts provided in the COSMOS database \citep{laigle2016cosmoscat}.

The COSMOS HSC transient survey has led to the discovery of nearly ten high-redshift SLSN candidates. Among them, we have thus far successfully confirmed redshifts in three high-redshift SLSNe \citep{curtin2018slsn}. The identified redshifts are $z = 2.399$, $z = 1.965$, and $z = 1.851$. Unfortunately, the spectra are not good enough to identify spectroscopic type. There are several candidates with higher host photometric redshifts, $z\sim3.2$ and $z\sim4.2$, whose spectra have not been obtained \citep{moriya2018slsn}.

Based on the three SLSNe at $z\sim2$, \citet{moriya2018slsn} estimated SLSN rates at $z\sim 2$ of $\sim 900~\mathrm{Gpc^{-3}~yr^{-1}}$. This rate, based solely on spectroscopically confirmed SLSNe, is already comparable to the total SLSN rate at $z\sim2$, estimated by extrapolating the local SLSN rate \citep{quimby2013slsnrate} using the cosmic star-formation history \citep{madau2014sfr}. The SLSN rate at $z\sim 2$ may be higher if we take SLSN candidates without spectroscopic confirmation into account.

\section{Summary}

We performed a deep transient survey with HSC of the Subaru telescope. The ultra-deep layer, the central 1.77 deg$^2$ of the COSMOS field with one HSC pointing, was observed repeatedly for 6 months in 2016 and 2017 with $g$, $r$, $i$, $z$, and $y$ filters, while the deep layer, 5.78 deg$^2$ with four HSC pointings, was observed for 4 months. For each month, data were taken at two epochs per filter. The limiting magnitudes per epoch for the ultra-deep layer are as follows: $26.4$, $26.3$, $26.0$, $25.6$, and $24.6$ mag in the $g$-, $r$-, $i$-, $z$-, and $y$-band, respectively; Deep-layer values are roughly 0.6 mag shallower. The data set obtained is one of the deepest wide-field transient surveys attempted to date.

In the dataset, 1,824 SN candidates were identified. Among our samples, 207 and 371 objects have spec-z and COSMOS photo-z , respectively. The median redshift is $z = 0.85$, and 187 objects (32\%) are located at $z > 1$. This is among the largest high-redshift SN samples. By light curve fitting using SALT2, 433 (129 with spec-z or COSMOS photo-z) candidates were classified as ``SNe Ia''. In particular, 58 objects are located at a redshift beyond $z = 1$. More dedicated photometric classification will be presented in a forthcoming paper.

Our dataset doubles the number of Type Ia SNe candidates at $z > 1$ for Type Ia SN cosmology, and the great depth enables a search for SLSNe at even higher redshifts \citep{moriya2018slsn,curtin2018slsn}. The survey also provides Type IIP SNe at medium redshift, which has been demonstrated by the highest-redshift Type IIP SNe for cosmological use  \citep{dejaeger2017}. In addition to the transient science, deep time-series images can also be used for studies of variable stars and AGN. This survey of the COSMOS field is the first half of the HSC-SSP transient survey. A similar transient survey for the SXDS field will be conducted over a period of 6 months, starting in September 2019. More details regarding the science topics to be covered, as well as the results from the forthcoming SXDS survey, will be presented in separate papers.

\begin{ack}
The Hyper Suprime-Cam (HSC) collaboration includes the astronomical communities of Japan, Taiwan, and Princeton University. The HSC instrumentation and software were developed by the National Astronomical Observatory of Japan (NAOJ), the Kavli Institute for the Physics and Mathematics of the Universe (Kavli IPMU), the University of Tokyo, the High Energy Accelerator Research Organization (KEK), the Academia Sinica Institute for Astronomy and Astrophysics in Taiwan (ASIAA), and Princeton University.  Funding was contributed by the FIRST program from the Japanese Cabinet Office, the Ministry of Education, Culture, Sports, Science and Technology (MEXT), the Japan Society for the Promotion of Science (JSPS),  the Japan Science and Technology Agency  (JST),  the Toray Science  Foundation, NAOJ, Kavli IPMU, KEK, ASIAA, and Princeton University.

The Pan-STARRS1 Surveys (PS1) have been made possible through contributions by the Institute for Astronomy, the University of Hawaii, the Pan-STARRS Project Office, the Max-Planck Society and its participating institutes, the Max Planck Institute for Astronomy, Heidelberg and the Max Planck Institute for Extraterrestrial Physics, Garching, Johns Hopkins University, Durham University, the University of Edinburgh, Queen's University Belfast, the Harvard-Smithsonian Center for Astrophysics, the Las Cumbres Observatory Global Telescope Network Incorporated, the National Central University of Taiwan, the Space Telescope Science Institute, the National Aeronautics and Space Administration (under Grant No. NNX08AR22G issued through the Planetary Science Division of the NASA Science Mission Directorate), the National Science Foundation (under Grant No. AST-1238877), the University of Maryland, and Eotvos Lorand University (ELTE).
 
This study used software developed for the Large Synoptic Survey Telescope (LSST). We thank the LSST Project for making their code available as free software at http://dm.lsst.org.
 
IT and NY acknowledge financial support from JST CREST (JPMHCR1414).
MT is supported by an Inoue Science Research Award from the Inoue Foundation for Science and the Grant-in-Aid for Scientific Research programs of JSPS (15H02075, 16H02183) and MEXT (17H06363).

This research is based in part on data collected at the Subaru Telescope and retrieved from the HSC data archive system, which is operated by the Subaru Telescope and Astronomy Data Center at NAOJ.
\end{ack}

\bibliographystyle{myaasjournal}
\bibliography{hsc}

\end{document}